\documentclass[10.5pt,a4paper,notitlepage,superscriptaddress,twocolumn,showkeys,pre, aps, floatfix,nofootinbib]{revtex4-1}
\usepackage[scale=0.8]{geometry}
\usepackage[utf8]{inputenc}
\usepackage[english]{babel}
\usepackage{amssymb}
\usepackage{amsmath}
\usepackage{graphicx}
\usepackage[caption=false]{subfig}
\usepackage{float}
\usepackage[normalem]{ulem}
\usepackage{color}
\usepackage{placeins}
\usepackage{dashbox}
\usepackage{hyperref}
\usepackage[bottom]{footmisc}
\usepackage{cleveref}
\usepackage{movie15}
\usepackage{blindtext}

\hypersetup{
    colorlinks=true,
    linkcolor=black,
    filecolor=black,  
    citecolor=black,    
    urlcolor=black,
    linktoc=all,
} 
\newcommand{\av}[1]{\langle #1 \rangle}

\begin{document}

\author{Elisenda Ortiz}
\affiliation{Departament de F\'{i}sica de la Mat\`{e}ria Condensada, Universitat de Barcelona, Mart\'{i} i Franqu\`{e}s 1, 08028 Barcelona, Spain}
\affiliation{Universitat de Barcelona Institute of Complex Systems (UBICS), Universitat de Barcelona, Barcelona, Spain}

\author{Guillermo Garc\'{i}a-P\'{e}rez}
\affiliation{QTF Centre of Excellence, Turku Centre for Quantum Physics, Department of Physics and Astronomy, University of Turku, FI-20014 Turun Yliopisto, Finland}
\affiliation{Complex Systems Research Group, Department of Mathematics and Statistics, University of Turku, FI-20014 Turun Yliopisto, Finland}

\author{M. \'{A}ngeles Serrano}
\email{marian.serrano@ub.edu}
\affiliation{Departament de F\'{i}sica de la Mat\`{e}ria Condensada, Universitat de Barcelona, Mart\'{i} i Franqu\`{e}s 1, 08028 Barcelona, Spain}
\affiliation{Universitat de Barcelona Institute of Complex Systems (UBICS), Universitat de Barcelona, Barcelona, Spain}
\affiliation{ICREA, Pg. Llu\'{i}s Companys 23, E-08010 Barcelona, Spain}

\date{\today}

\title{Geometric detection of hierarchical backbones in real networks}

\begin{abstract}
Hierarchies permeate the structure of real networks, whose nodes can be ranked according to different features. However, networks are far from tree-like structures and the detection of hierarchical ordering remains a challenge, hindered by the small-world property and the presence of a large number of cycles, in particular clustering. Here, we use geometric representations of undirected networks to achieve an enriched interpretation of hierarchy that integrates features defining popularity of nodes and similarity between them, such that the more similar a node is to a less popular neighbor the higher the hierarchical load of the relationship. The geometric approach allows us to measure the local contribution of nodes and links to the hierarchy within a unified framework. Additionally, we propose a link filtering method, the \textit{similarity filter}, able to extract hierarchical backbones containing the links that represent statistically significant deviations with respect to the maximum entropy null model for geometric heterogeneous networks. We applied our geometric approach to the detection of similarity backbones of real networks in different domains and found that the backbones preserve local topological features at all scales. Interestingly, we also found that similarity backbones favor cooperation in evolutionary dynamics modelling social dilemmas.
\end{abstract}\

\maketitle

Many real systems display a hierarchical organization~\cite{Corominas2013} where higher status members dominate over lower-graded ones, according to a certain measure of power, wealth, importance or influence. Examples are ubiquitous in living systems, including molecular regulators governing gene expression~\cite{Erwin2009}, animal communities of eusocial insects~\cite{Johnson2014}, dominant-subordinate relationships in mammals~\cite{Samuels1984}, and different structures ---companies, political parties, courts, military, organized religion etc---in human society~\cite{saunders1990}. Additionally, hierarchical organization can be found in non-living systems such as computer generated imagery (CGI)~\cite{deloura2001game}, grammatical theory of language~\cite{allerton1979} or the structure of a musical composition~\cite{yeston1976}. Hierarchies are, thus, ubiquitous, and shape more easily controllable structures~\cite{mones2012hierarchy} that can emerge as the result of opposing forces, such as competition between individuals~\cite{theraulaz1995self} or a combination of cooperation and imitation strategies~\cite{eguiluz2005cooperation}. \\

Complex networked systems, however, present important challenges when it comes to the detection of hierarchies due to the lack of a unique and unambiguous stratification scheme, possibly including nestedness or layered structure. This is caused by the small world property and the presence of a large number of cycles of different lengths in their topologies, in particular clustering~\cite{Ravasz2003}, so that networks' organization deviates strongly from tree-like. In directed networks, link directionality  can be exploited to ease the problem and hierarchical order can be detected using, for instance, penalty-function minimization strategies~\cite{Costa2004,Nikolaj2015,Letizia2018, Nikolaj2019}. Nevertheless, most frequently the only meaningful or available representation of a complex system is an undirected graph. Within this architecture, a hierarchy is typically defined as a ranking where the status of a node becomes determined by some heterogeneous topological property, for instance degree~\cite{Trusina2004} or some other centrality measure~\cite{perra2008spectral,mones2012hierarchy}. However, other attributes shape as well the hierarchical structure of real networks, like similarity between nodes ~\cite{Serrano:2008ga}. Clearly, the control exerted by a higher-status node over a lower-status one will be more effective when there exists closeness or affinity between them. Conversely, the strength of hierarchical relations gets dissolved as nodes loose their proximity and become dissimilar.

Here, we integrate degree rank, or popularity, and similarity between nodes in an enriched interpretation of hierarchy, valid for real networks. For this purpose, we capture network architecture using geometric network maps~\cite{Boguna2010,Garcia2019} and models~\cite{Serrano:2008ga,KrPa10}, which naturally encode popularity and similarity attributes of nodes as coordinates in an underlying metric space. Exploiting the geometric approach, we are able to characterize the individual contribution of each node and each link to the hierarchical structure of a network. Moreover, we exploit the great heterogeneity found in the hierarchy load of links to propose a filtering method, the similarity filter, that offers a practical procedure to extract a hierarchical similarity backbone of a network. The obtained similarity backbones contain the links that represent statistically relevant interactions with respect to the maximally random geometric network organization, constrained by node degrees and clustering~\cite{Boguna:2020}. The similarity filter preserves network features at all scales while pruning a large number of links, in the spirit of the disparity filter for weighted networks~\cite{Serrano:2009a}. However, the similarity filter has a different purpose and operates on the basis of different models in a completely distinct framework, that of unweighted undirected networks. To illustrate the use and results of the similarity filter, we extracted and analyzed the hierarchical similarity backbones of several real-world networks from different domains. Finally, we explored the role of hierarchical similarity backbones in evolutionary dynamical processes, which historically have been argued to be sensitive to the hierarchical organization of complex architectures~\cite{Simon1962,Agre2003}. Implementing an evolutionary prisoner's dilemma game~\cite{Szabo2007,Szabo1998,Nowak2006,Weibull1995} on the real networks under study, we discovered that the similarity backbones tend to achieve final states of greater cooperation as compared to the corresponding original networks, when initial conditions are equivalent.

\section{Hierarchy load of links and nodes}

We base our definitions of hierarchy on geometric maps of real networks obtained from geometric network models~\cite{Serrano:2008ga,KrPa10}. In these models, the high clustering typically observed in real networks emerges naturally as a result of the metric properties of an underlying space, which abstracts the similarity space among nodes. In this paper, we use the $\mathcal{S}^1$ model, in which the similarity space is a one-dimensional sphere (a circle). Nodes thus have an angular coordinate $\theta_i$, so that the more similar two nodes are the shorter is the angular separation, $\Delta \theta_{ij}=\mathrm{min}(|\theta_i - \theta_j|,2\pi-|\theta_i-\theta_j|)$, between them. Nodes are also characterized by a hidden degree $\kappa_i$, directly related to the observed degree, standing for popularity or status. The probability that two nodes $i$ and $j$ connect  is a function that increases with their similarity and with the product of their hidden degrees 
\begin{equation}
 \label{eq:probability_of_connection_S1}
  p_{ij} = \frac{1}{1 + \left( \frac{R \Delta\theta_{ij}}{\mu \kappa_i \kappa_j} \right)^\beta} \ ,
\end{equation}
where $R$ is the radius of the similarity circle, which we choose to be equal to $N/(2 \pi)$ for $N$ nodes, so that the density of nodes in the circle is equal to one, and parameters $\beta$ and $\mu$ control the mean clustering coefficient and the average degree of the network, respectively.

The $\mathcal{S}^1$ model is able to reproduce many of the features widely observed in real complex networks, such as scale-freeness, high levels of clustering and the small-world property, among others~\cite{Serrano:2008ga}. Interestingly, the $\mathcal{S}^1$ model is the only model able to produce maximum entropy ensembles with power-law degree distributions and clustering and without nonstructural degree correlations~\cite{Boguna:2020}. Moreover, the model also allows us to construct geometric maps of real networks through a process called \textit{network embedding}. Given a real network, one can find the values $\lbrace \kappa_i,\theta_i \rbrace$ for every node $i$, as well as the global parameters $\mu$ and $\beta$, that maximize the likelihood for the observed network to be generated by the model~\cite{Boguna2010}. The congruency of the $\mathcal{S}^1$ model with real networks results in very meaningful embeddings, which have proven to be useful for network navigation~\cite{Boguna2010,gulyas2015navigable,Elis2017}, and for the discovery of symmetries in the structure of real networks such as self-similarity at different length scales~\cite{Garcia2018Renorm}. Notice that the distribution of angular positions of nodes in the $\mathcal{S}^1$ model is assumed to be homogeneous, while typically real network embeddings show that nodes form geometric communities in similarity space~\cite{BoPa10,Serrano2012,Garcia2016Atlas,allard2020navigable}. The geometric embeddings of real networks used in this work were computed using the mapping tool Mercator~\cite{Garcia2019}. 

We use the purely geometric isomorphic version of the $\mathcal{S}^1$ model, named the $\mathcal{H}^2$ model~\cite{KrPa10}, for visualization purposes. In the $\mathcal{H}^2$ model the hidden degree is transformed into a radial coordinate in a hyperbolic two-dimensional disk, such that higher degree nodes are placed closer to the centre of the disk. The angular coordinate remains as in the $\mathcal{S}^1$ circle and the connection probability becomes a decreasing function of the hyperbolic distance between nodes, see Fig.~1(a). For further details refer to section Methods B.

\subsection{Definition of hierarchy load}

We characterize the local contribution of a link or a node to the hierarchical structure of a network by measuring its \textit{hierarchy load}, which depends on status, similarity, and the reference provided by a null model to discount the effects of random fluctuations. We use the $\mathcal{S}^1$ as a null model since it is the maximum entropy model for geometric networks with heterogeneous degrees~\cite{Boguna:2020}. It provides expectations for the distribution of hierarchy loads in a pure random assignment of angular positions of nodes given a degree distribution and a level of clustering, so that anomalous fluctuations can be detected.\\

\begin{figure*}[t]
\centering
\scalebox{1.0}{
  \includegraphics[width=1.0\linewidth]{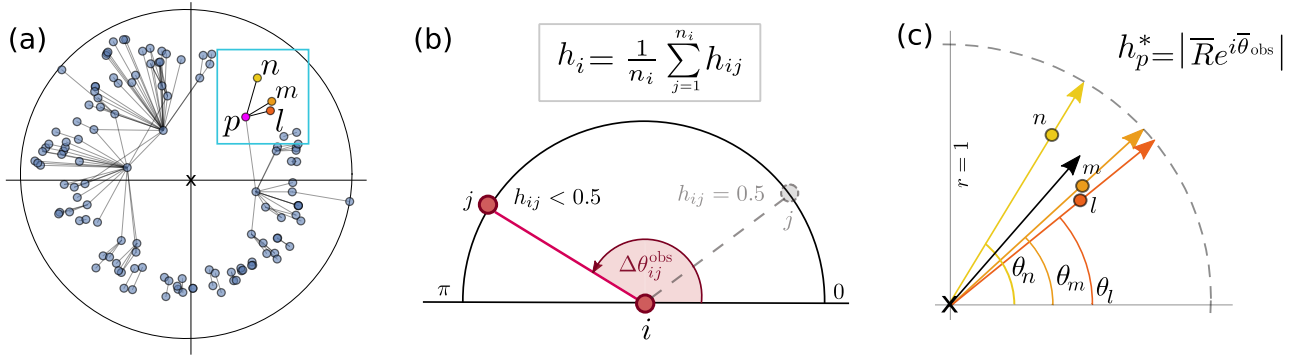}
}
	\caption{\textbf{Hierarchy load in geometric networks.} \textbf{(a)} Illustration of a geometric network in the hyperbolic disk. Node $p$ and its subordinates $l,m,n$ are highlighted inside a blue square. Notice node $p$ has a fourth neighbor which is not a subordinate since it lies at a smaller radial position from the center of the disk and thus has higher degree than node $p$. \textbf{(b)} Hierarchy load of a link, $h_{ij}$, between a generic node $i$ and a lower-status neighbor $j$ located at angular distance $\Delta\theta_{ij}^{\mathrm{obs}}$. Notice in this example $h_{ij}$ is necessarily $h_{ij}<0.5$ as $j$ is located at an angular distance greater than the expected from the null model, which appears depicted in grey. The hierarchy load of node $i$, $h_i$, is obtained from averaging all link hierarchy loads corresponding to this node. \textbf{(c)} Hierarchy load of node $p$ from pannel A as the circular mean of the angular coordinates of lower-status neighbors $l,m,n$. Mean resultant vector is depicted by a black arrow of $|\overline{R}|<1$. Vectors added in the calculation of $h_p^*$ (see Eq.\ref{Eq.2}) appear as arrows in yellow, orange and red inside a circle of unit radius.}
      \label{Fig.2Sketch_similarity_filter}
\end{figure*}

Given a geometric network embedding, where a node $i$ has coordinates $\lbrace \kappa^{\mathrm{obs}}_i,\theta^{\mathrm{obs}}_i \rbrace$, we consider that popularity, corresponding to the hidden degree or, equivalently, to the radial position in the hyperbolic plane, is a measure of status. This means that a node $i$ has a lower-status neighbor $j$ when $\kappa_j^{\mathrm{obs}} < \kappa_i^{\mathrm{obs}}$. Similarity between the two nodes is represented by their angular separation in the network map, $\Delta\theta_{ij}^{\mathrm{obs}} = \min \left( | \theta_i^{\mathrm{obs}} - \theta_j^{\mathrm{obs}} |, 2 \pi - | \theta_i^{\mathrm{obs}} - \theta_j^{\mathrm{obs}} | \right)$. \\

First, we define the hierarchy load $h_{ij}$ of a link between node $i$ and its lower-status neighbor $j$ as the probability of obtaining a similarity distance between them in the $\mathcal{S}^1$ model greater than the one observed in the map
\begin{equation}
h_{ij}={P(\Delta\theta_{ij} > \Delta\theta_{ij}^{\mathrm{obs}}}).
\label{Eq.2}
\end{equation}
The rationale behind this definition is that a high probability in the null model for the angular separation between nodes $i$ and $j$ to be larger than observed indicates that they are closer than expected in similarity space, hence signaling a highly hierarchical connection.
 Being a probability, $h_{ij}$ is bounded in the interval [0,1], while the angular separation between nodes $\Delta\theta_{ij}^{\mathrm{obs}}\in [0,\pi]$. Furthermore, Eq.~\eqref{Eq.2} can be computed analytically giving an expression depending on the coordinates of the nodes in the embedding (see Methods D). In particular, in a synthetic network generated with the $\mathcal{S}^1$ model, the expected value of $h_{ij}$ for any $\lbrace i,j\rbrace$ is $h_{ij} = 1/2$ since, in that case, the observed angular distance $\Delta\theta_{ij}^{\mathrm{obs}}$ is generated by the model and, hence, it is a random variable distributed according to the same distribution as $\Delta\theta_{ij}$. In other words, Eq.~\eqref{Eq.2} reduces to the probability for two equally distributed variables $a$ and $b$ to fulfill $a > b$. As a consequence, Eq.~\eqref{Eq.2} has the clear advantage of being size independent, in the sense that the value $h_{ij} = 1/2$ defines the $\mathcal{S}^1$-model hierarchy baseline for any $N$, therefore allowing us to compare the hierarchy structure of networks of different size. 

Link hierarchy loads inform about how substantial is the contribution of a link towards the hierarchy by comparison with the reference level provided by the $\mathcal{S}^1$ null model, which is $h_{ij}=1/2$ as explained above. Accordingly, when the link hierarchy load is higher than the reference, $h_{ij} > 1/2$, nodes $i$ and $j$ are closer than expected in angular distance, and so they are more similar and their relationship is more hierarchical. In contrast, when the probability is low, $h_{ij} < 1/2$, $i$ and $j$ are more dissimilar than expected by the null model, meaning that they lie farther away in the angular space and, thus, their relationship is less hierarchical. In fact, $h_{ij}=0$ if $\Delta\theta_{ij}^{\mathrm{obs}}=\pi$, while $h_{ij}=1$ if $\Delta\theta_{ij}^{\mathrm{obs}}=0$.\\

Finally, within the same framework we also define a measure of hierarchy load for nodes as 
\begin{equation}
h_i = \frac{1}{n_i}\sum_{j=1}^{n_i} h_{ij}, 
\label{Eq.hi}
\end{equation}
where the sum runs over the $n_i$ lower-status neighbours $j$ of node $i$ satisfying $\kappa_j < \kappa_i$.

\subsection{Hierarchy load of nodes in terms of angular concentration}
Alternatively, one can also measure the hierarchy load of a node $i$ as the angular concentration of its $n_i$ lower-status neighbors by computing the \textit{circular mean} of their angles $\lbrace \theta_{j} \rbrace_{j=1,...,n_i}$. This method is different but very similar to the node hierarchy measure used in~\cite{Garcia2016Atlas} for the analysis of the World Trade Web. When lower-status neighbors are concentrated in a narrow angular sector it means that node $i$ has a tendency to establish links with more similar nodes, hence node $i$ is contributing towards a more hierarchical structure and thus carrying a higher hierarchy load $h_i$. Conversely, when the lower-status neighbors of $i$ are distributed in a very broad angular sector, the hierarchy load of node $i$ is low, indicating that it is able to establish links with more dissimilar lower-status nodes, thus supporting a flatter organization. 

The hierarchy load of a node in terms of angular concentration of its $n$ lower-status neighbors is computed as the length of the mean resultant vector,
\begin{equation}
h_{\color{blue}}^* =\left\vert\frac{1}{n}\sum_{j=1}^{n}e^{i\theta_j^{\mathrm{obs}}} \right\vert= \left\vert \overline{R}e^{i\overline{\theta}{\mathrm{obs}}}\right\vert=\overline{R},
\label{Eq.1}
\end{equation}
see Fig.~1(c). The modulus of the circular mean vector $\overline{R} \in [0,1]$ is a good proxy of angular confinement since it is 0 for angles pointing in opposite directions and it becomes 1 when the angles are totally aligned. The measure is simple enough so that it generalizes well to networks of very different domains as long as they admit a geometric interpretation. It is worth mentioning that, since the average angular separation between nodes decreases with the network size $N$, this quantity increases with network size. This should be taken into account when comparing $h^*$ measurements among networks of different size.

\begin{figure}[t]
\centering
  \includegraphics[width=1\linewidth]{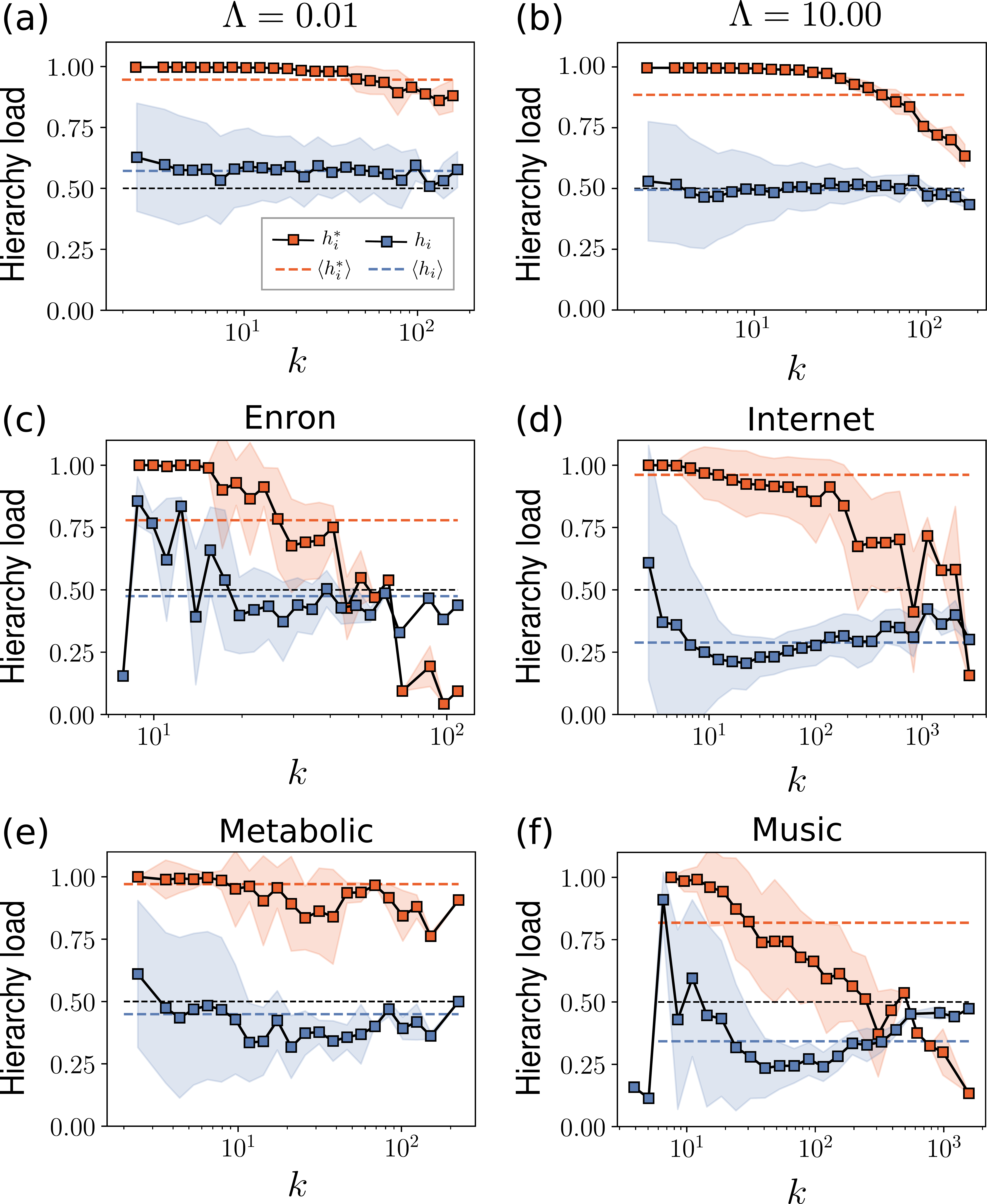}
  \caption{\textbf{Hierarchy load spectra of synthetic and real networks.} \textbf{(a-b)} Hierarchy load spectrums for synthetic SCSS networks of size $N=1000$ nodes, generated with power-law degree distribution exponent $\gamma= 2.50$, clustering parameter $\beta=2.50$ and variable attractiveness $\Lambda$. Results are averaged over 10 network realizations for each choice of $\Lambda$. \textbf{(c-f)} Hierarchy load spectra of 4 real networks. In all plots blue curves correspond to $h_i$ in Eq.~\eqref{Eq.hi} computed from the link hierarchy loads while red curves correspond to $h_i^*$ in Eq.~\eqref{Eq.1} as given by the circular mean resultant vector. Dashed blue and dashed red lines indicate the corresponding average hierarchy load values for the whole network. Dashed black lines provide the reference level of the model, placed at hierarchy load $1/2$. }
  \label{Fig.2}
\end{figure}

\subsection{Hierarchy load vs geometric communities}
To understand how the global distribution of angles, and, in particular, the existence of geometric communities, could affect the spectrum of hierarchy loads, we consider synthetic networks with controllable angular concentration of nodes. Typically, real network maps present angular regions more densely populated than others, which define a partition of the network into different geometric groups~\cite{BoPa10,Serrano2012,Garcia2016Atlas,allard2020navigable}. Geometric communities can be accurately detected using algorithms such as the Topological Critical Gap Method~\cite{Garcia2016Atlas} and reproduced by geometric network models~\cite{Garcia2018,Zuev2015}. We generated synthetic networks with tunable geometric communities using the soft communities in similarity space (SCSS) model~\cite{Garcia2018} (see Methods B). The SCSS model is an extension of the $\mathcal{S}^1$ model that enables to create scale-free networks with high clustering while controlling for the global heterogeneity of the distribution of angular coordinates. Essentially, this model relies on a preferential attachment process in similarity space~\cite{Zuev2015}, so that the angular coordinates of nodes depend on the angular coordinates of higher-degree nodes. The effect of this similarity preferential attachment is regulated by a parameter $\Lambda$. The SCSS recovers the $\mathcal{S}^1$ model in the limit of homogeneous angular distributions, which corresponds to $\Lambda \to \infty$.

We show the spectra of hierarchy loads $h_i$, Eq.~(\ref{Eq.hi}), and $h_i^*$, Eq.~(\ref{Eq.1}), in Fig.~2(a-b) for synthetic networks with very different geometric community strengths ($\Lambda=0.01$ and $\Lambda=10.0$). The spectra of hierarchy loads is measured by averaging the node hierarchy loads over degree classes. The two measures display different results, but in both cases the global angular heterogeneity has a minor effect in shaping the hierarchy loads of nodes. Therefore, the spectrum of hierarchy loads of nodes is not merely a measurement of geometric community structure. As expected, $h_i$ spectra are flat and lie around the average hierarchy load $\av h= \sum^N_{i=1}h_i/N$ for the whole network. At large $\Lambda$ values, $\av h$ tends to $0.5$ because, by construction, SCSS networks recover the $\mathcal{S}^1$ model in this limit. Heterogeneous angular distributions in the limit of small $\Lambda$ values reduce the average angular distance between nodes in the network, and as a consequence the average hierarchy load of the network is slightly above $0.5$.

This effect is also evident in the spectra of hierarchy loads $h_i^*$, where $\av h^*$ is lower for more homogeneous angular distributions ($\Lambda=10.00$, $\av h^*= 0.89 \pm 0.10$), as compared with networks with more heterogeneous distributions that present higher values ($\Lambda=0.01$, $\av h^* = 0.95 \pm 0.06$). In this case, the average hierarchy level of the synthetic networks is rather high ($\av h ^* \approx 0.9$), basically due to sustained large values of hierarchy load for a wide range of node degrees.

\label{Synthetic & Real}
\subsection{Hierarchy spectrum of real networks}

We measured and compared the hierarchy spectrum of several real networks from different domains: the email communication network within the Enron company (Enron)~\cite{Klimt2004}, the Internet at the autonomous system level (Internet)~ \cite{Boguna2010,Claffy2009}, the one-mode projection onto metabolites of the human metabolic network at the cell level (Metabolic)~\cite{Serrano2012} and  the network of chord transitions in western popular music (Music)~\cite{Serra2012}. For more information see Table~\ref{Tab:Real nets properties} and Methods~A.

Real networks show a variety of profiles, see blue curves in Fig.~2(c-f). They also present in all cases an average hierarchy load $\av h$ below the reference of $1/2$. In particular, Enron shows the highest average hierarchy load at the node level, $ \av h= 0.47 \pm 0.19 $, followed by Metabolic, Music and, lastly the Internet with $\av h = 0.29 \pm 0.20 $. Variations in $\av h$ across networks conform to their distinct spectra. For instance, whereas Music and Internet networks show noticeable fluctuations in $h_i$ across degree classes, these are milder for Enron and Metabolic networks. In general, however, all networks show a tendency for the lowest degree nodes to have the highest hierarchy loads and for the highest degree nodes, or hubs, to approach $h_i = 1/2$. This last observation may be attributable to great heterogeneity in the hierarchy load of links $h_{ij}$ and the fact that hubs present numerous connections with lower status nodes which are averaged when computing $h_i$.

Regarding the hierarchy load of nodes in terms of angular concentration, the Internet and Metabolic are significantly more hierarchical ($\av h^* \approx 1$) than Music and Enron. The particular trend followed by each profile  has its roots in each network's specific degree sequence, whereas angular communities show again minor influence as revealed by the resemblance between the spectra of the 4 real networks with the spectra of their corresponding geometrically randomized counterparts (see Methods~B and Appendix Fig.~\ref{Fig.6}). The randomized counterparts consist in replicas of the real networks where the distribution of angular coordinates is homogeneized, thus eliminating geometric community structure, while the rest of properties are preserved and the replica network remains maximally geometric. Moreover, we note that the hierarchy loads of nodes tend to show strong heterogeneity and an inverse correlation with the degree, so that more popular nodes (higher $k$) contribute more to dilute the hierarchical structure by connecting to less affine lower-popularity neighbors. This trend is remarkably pronounced for the Internet, Enron, and Music, while heterogeneity is less pronounced in Metabolic. In fact, this happens because the Metabolic network presents more modular hubs, less prone to connect with dissimilar nodes due to an unusually marked partition of the similarity space.

\label{The similarity filter}
\section{The similarity filter and hierarchical backbones of real networks}

\begin{figure}[t]
\centering
  \includegraphics[width=0.70\linewidth]{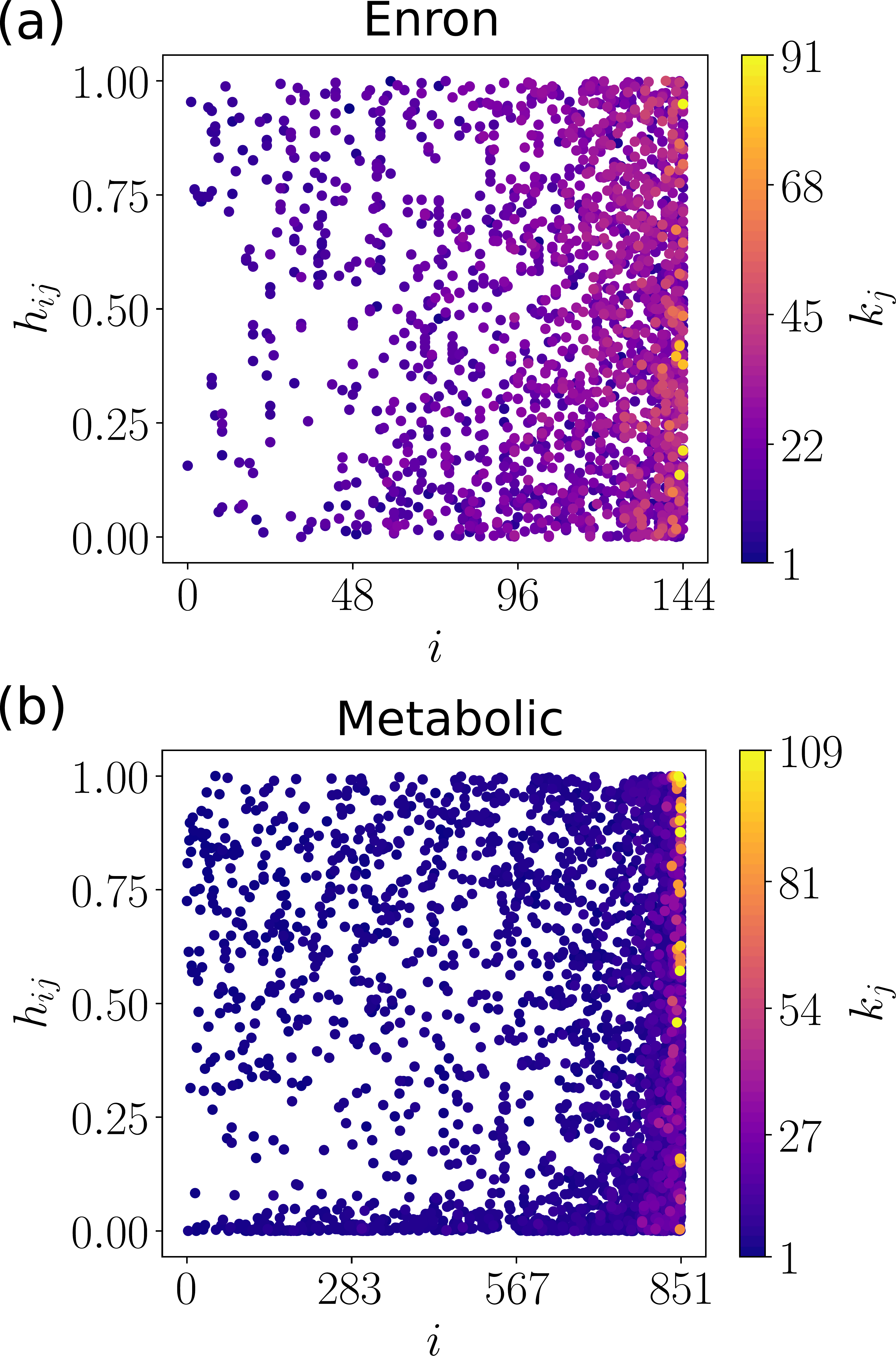}
	\caption{\textbf{Link hierarchy loads of real networks.} \textbf{(a-b)} Link hierarchy loads of the Enron and Metabolic networks, respectively. Each dot indicates the value of $h_{ij}$ of a link stablished between node $i$ and one of its lower-status neighbors $j$ with hidden degree $\kappa_j$ indicated by the color code. Notice node labels in the x-axis are sorted by increasing hidden degree $\kappa_i$, so that more popular $i$ nodes appear to the right.}
      \label{Fig.3Heterogeneity}
\end{figure}

Strong heterogeneity is found in real networks if we analyze the contribution of lower-popularity neighbors to the hierarchy load of nodes, see Fig.~3. The link hierarchy load contributions involving low popularity neighbors are represented in Fig.~3 by dots color-coded to the blue end of the scale. We find those dots covering almost all the area of the plots, thus indicating that these link hierarchy loads span the entire range of $h_{ij}$, both when the link is shared with another node of low degree or with a node of high degree (right end of the x-axis). This observation reveals that the vast majority of link hierarchy loads in the network are distributed rather heterogeneously. We can take advantage of this heterogeneity to filter out the connections that dominate the hierarchical organization of the network in terms of popularity and similarity, what we name the hierarchical similarity backbone (HSB). A hierarchical similarity backbone contains the links that represent statistically significant contributions with respect to the null hypothesis given by the $\mathcal{S}^1$ model, that is the only model able to produce maximum entropy ensembles constrained by power-law degree distributions and clustering and without nonstructural degree correlations~\cite{Boguna:2020}.

\begin{figure*}[t]
\centering
\scalebox{1.0}{
  \includegraphics[width=1.0\linewidth]{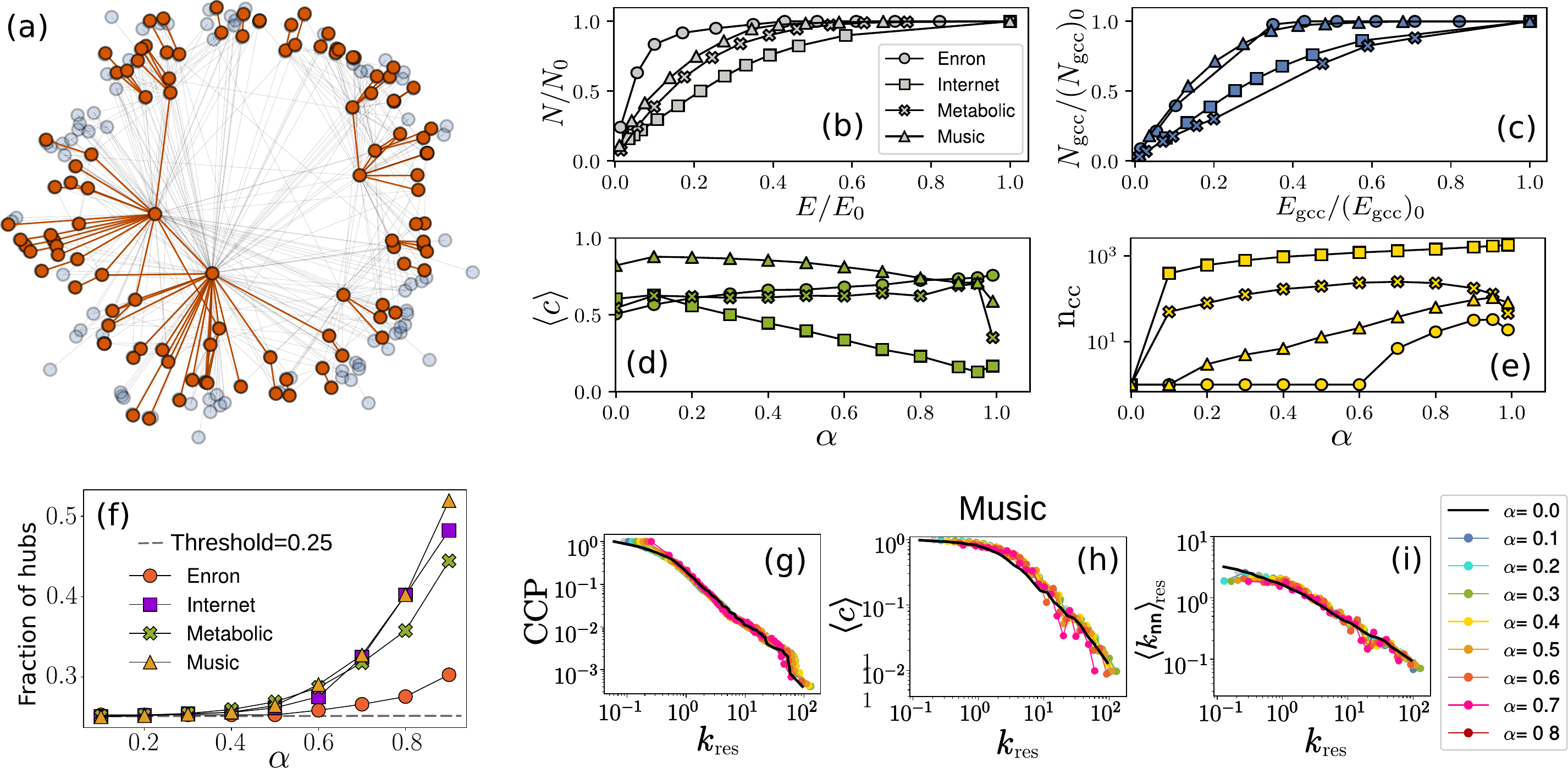}
}
	\caption{\textbf{Hierarchical similarity backbones.} \textbf{(a)} A hierarchical similarity backbone of the World Trade Web~ \cite{Garcia2016Atlas} (in red) filtered with $\alpha=0.4$, on top of the complete network (pale blue). \textbf{(b-e)} Plots show in order: the relative number of nodes in the backbone, against the relative number of edges in the backbone (subindex $0$ refers to the complete network); the relative number of nodes in the backbone giant connected component ($gcc$), against the relative number of edges in the backbone $gcc$; mean clustering coefficient of the backbone for increasing $\alpha$; number of disconnected components against increasing values of the significance $\alpha$ \textbf{(f)} Fraction of nodes considered hubs, for threshold value $\tau=0.25$, found in backbones obtained with increasing $\alpha$. \textbf{(g-i)} Topological features of the HSBs of the Music network, obtained for $\alpha$'s from 0.1 to 0.8. Value $\alpha=0.0$ corresponds to the original unaltered network, whereas in the most restrictive HSB ($\alpha=0.8$) $0.60\%$ of the nodes and $14\%$ of the links remain. From left to right: complementary cummulative probability distribution (CCP) of rescaled degrees, $k_{\mathrm{res}} = k/\langle k\rangle $, degree dependent clustering coefficient $\av c$ over rescaled-degree classes, normalised average nearest neighbor degree $\langle k_{\mathrm{nn}} \rangle_{\mathrm{res}}= \langle k_{\mathrm{nn}}(k_{\mathrm{res}})\rangle\langle k \rangle /\langle  k^2 \rangle$.}
      \label{Fig.HBBS_big_plot}
\end{figure*}

The link hierarchy load in Eq.(\ref{Eq.2}) measures the probability under the null hypothesis that the similarity distance between a node and a lower-status neighbor is larger than the observed in the embedding of the network, what is known as p-value in statistical inference. By imposing a significance level $\alpha$, the links that carry a hierarchy load that are not compatible with the random angular distribution of angles in the $\mathcal{S}^1$ model, and reject the null hypothesis, can be filtered out. A hierarchical similarity backbone is then obtained by preserving all the links that satisfy the criterion $h_{ij} \geq \alpha$, while discounting the rest. As we increase the significance level $\alpha \in [0,1]$, the filter progressively focuses on more relevant links to obtain a sequence of nested subgraphs, each with a more strict condition for a link to belong to the HSB of the network, see an illustration in Fig.~4(a).
Noteworthy, since the similarity filter is applied to the links, nodes of any degree may find a place in a very hierarchical backbone if they are found to have significantly strong hierarchical connections.

We tested the performance of the similarity filter by exploring the hierarchical similarity backbones of the four real networks considered in this work. Figure~4(b) shows that, for all real networks, low values of $\alpha$ reduce the number of links drastically while most of the nodes are preserved in the backbone. Notice that $\alpha$ increases from right to left in Figs.~4(b-c). For instance, when filtered with $\alpha =0.25$, the Internet, Metabolic and Music HSBs contain a proportion of edges that is already less than half of the original, so $E/E_0 < 0.5$. In contrast, the proportion of nodes in the same backbones remains very high, the lowest case being the Internet, but with still $85\%$ of the nodes. The results in Fig.~4(c) show similar behaviour for the reduction in nodes and edges of the giant connected components (\textit{gcc}'s) of the backbones, for the 4 networks under study. Only the decay in number of nodes in the $gcc$ of the backbones tends to be less abrupt than in Fig.~4(b) and start sooner, at smaller values of the filtering parameter $\alpha$. Moreover, we observe in Fig.~4(d) that the mean clustering coefficient of the filtered similarity backbones does not have strong fluctuations and varies little with $\alpha$, with the exception of the Internet which shows a clear decreasing trend.

In Fig.~4(f) we inspect the participation of hubs in the HSBs. For this purpose, we sort the nodes in the network from highest degree to lowest and tag as ``hubs'' all nodes lying within a top slice of the list, delimited by a threshold value $\tau$. For instance, when $\tau=0.25$, the top $25\%$ of the nodes in the ranked list are considered as hubs. Subsequently, we keep track of the proportion of such high-degree nodes in every hierarchical backbone for increasing $\alpha$. Figure 4(f) demonstrates that, even when considering that a large fraction of the network ($\tau=0.25$) are nodes of \textit{high} degree, in fact these nodes only represent, at best, half of the backbone composition (see Internet, Metabolic and Music in Fig.~4(f) for $\alpha=0.9$). In Appendix Fig.~\ref{Fig.9} we show analogous plots to Fig.~4(f) for a wider range of threshold values $\tau \in [0.05-0.30]$ providing further evidence that, while more restrictive HSBs become enriched with hubs, the similarity backbones still present an assorted composition in terms of node degrees.

Finally, we find that topological features ---degree distribution, clustering coefficient and average nearest neighbors degree--- of the original network are preserved in the subgraphs as we increase $\alpha$ and the backbone is progressively restricted to exceedingly hierarchical links, see Fig.~4(g-i) and Appendix Fig.~\ref{Fig.7}. Only for very high values of the significance level, $\alpha \gtrsim 0.8$, when not only the number of links but also the number of nodes is strongly reduced, the measured topological properties start to deviate from the original curves. This suggests some grade of self-similarity across the sequence of HSBs. 

\label{Game dynamics in HBBs}
\section{Game dynamics in HSBs}

As an illustration of the importance of HSBs, we study a dynamical process with intriguing behavior in networked systems: the evolution of cooperation. The evolution of cooperation has been studied  in different fields~\cite{Axelrod1984,Nowak2011,Higgs2015,Cant2010}, but it is not well understood yet in scale-free networks~\cite{Assenza2008,Santos2006,Szolnoki2008,Santos2006Evo,Gomez2007} or in real networks~\cite{Gracia2012PNAS}, which additionally present high levels of clustering and finite size effects. In fact, only recently some mechanisms have been proposed to aid cooperation in real networks~\cite{Kleineberg2017metric}, based precisely in the geometric approach followed in this contribution. Here, we show that, counterintuitively, our HSBs capture the links of real networks that better support cooperative behavior in evolutionary dynamics. We consider an evolutionary prisoner's dilemma game consisting of two players deciding to cooperate or defect with one another, and gaining a specific reward depending on which of the four possible outcomes takes place. The game proceeds in successive rounds; in each round every node accumulates a payoff $\pi_i$ resulting from playing the game with all its neighbors; after that, and before moving to the next round, the strategy (cooperate or defect) played by every node is updated simultaneously taking into account an imitation mechanism. The imitation step consist of each node $i$ deciding to adopt the strategy of a randomly chosen neighbor $j$ with a probability dependant upon the difference of their payoffs $(\pi_i-\pi_j)$, to reflect the tendency of copying more successful neighbors. See section C in Methods for more details.

\begin{figure*}[t]
\centering
  \includegraphics[width=1.0\linewidth]{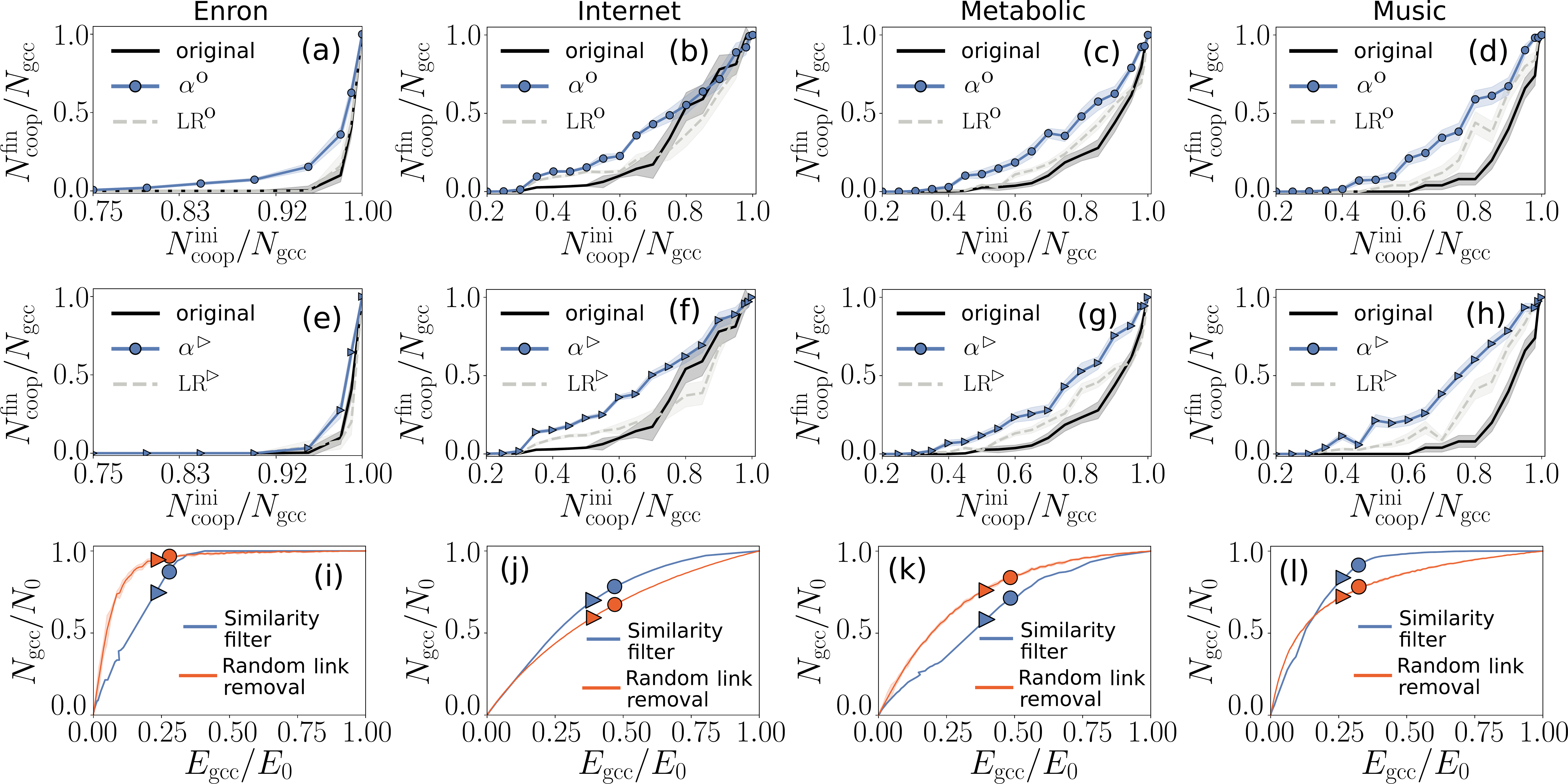}
  \caption{\textbf{Evolutionary game dynamics in HSBs of real networks.} \textbf{(a-h)} Fraction of final cooperators against fraction of initial cooperators for similarity backbones of Enron, Internet, Music and Metabolic networks. For every network we show two plots, each with the results of an HSB filtered with either $\alpha^{\mathrm{o}}$ or $\alpha^{\triangleright}$ (see numeric values in  Appendix Table~\ref{Tab:Backones_Evo_DYn}). For an HSB (blue curve with symbols), the corresponding random surrogate appears as a grey dashed line together with the original network in black line for reference. \textbf{(i-l)} Relative number of nodes against relative number of edges in the $gcc$ of similarity backbones, and their corresponding surrogates, for the four networks under study. Blue lines correspond to backbones obtained using the similarity filter and red lines to surrogates obtained by random link removal. Blue circle and triangle symbols highlight the fraction of nodes and edges of the 2 similarity backbones filtered with $\alpha^{\mathrm{o}}$ and $\alpha^{\triangleright}$, respectively. The same information is featured by red symbols for the random surrogates.}
\label{Fig.Dynamics}
\end{figure*}

We simulated the dynamics on the four real systems analyzed in this work and on two different HSBs for each of them (with $\alpha$ values and corresponding sizes in number of nodes and links reported in Appendix Table~\ref{Tab:Backones_Evo_DYn}). The results are provided in Fig.~5(a-h). Notice that the similarity backbones are always selected so that they face a considerable reduction in the number of links while their number of nodes does not decay drastically. That is, for a given real network, the HSBs where we run the game dynamics lie along the slope change part of the blue curves in Fig.~5(i-l), and are identified by blue symbols. We use random surrogates to discern whether the results of the dynamics on HSBs are due to their hierarchical nature. The surrogate backbones are obtained by removing a number of links at random so that the they have exactly the same number of links as the corresponding HSBs (see matching fraction of edges between red and blue symbols in Fig.~5(i-l)). As a result, the fraction of remaining nodes in a given HSB may be higher or lower than in the analogous random surrogate. Regardless of the situation, however, the dynamics results stay consistent, see red curves in Fig.~5(i-l). Note that the fraction of nodes in the random surrogates is still very high even when the number of links has been greatly reduced, akin to the case of similarity backbones. This was expected from the reported robustness of scale-free networks to random removals~\cite{PhysRevLett.85.4626,PhysRevLett.85.5468}.

The evolutionary dynamics are initiated by distributing a proportion of initial cooperators uniformly at random among the nodes of a network. For each of the three graphs (original network and the two similarity backbones), we vary the proportion of initial cooperators and quantify the level of cooperation achieved in the network at the end of the dynamics by measuring the fraction of final cooperators $N^{\mathrm{fin}}_{\mathrm{coop}}/N_{gcc}$ after $10^5$ rounds of the game. Notice no node alters its strategy while playing with its neighbors during an individual round (see Mehods C). The fraction of final cooperators $N^{\mathrm{fin}}_{\mathrm{coop}}/N_{gcc}$ is averaged over $100$ realizations in all showcased curves in Fig.~5 showing the results for the dynamics on HSBs. Notice the system can reach a quenched state before the maximum number of rounds is achieved if all agents become either cooperators or defectors. In that situation the imitation mechanism does not induce any further evolution and the dynamics becomes effectively frozen.\\

The results in Fig.~5(a-h) show that the real networks have a tendency for their hierarchical similarity backbones to display final cooperation levels equal or greater than the achieved in the original network for equal proportions of initial cooperators, despite their radically reduced number of links. We expect the dynamics not to work as well as they would in moving the system towards full cooperation when the number of agents to convince is still large but the nodes play the game within small groups of neighbors, as it happens in the backbones. This is because the probability of adopting a neighbor's strategy, $P_{i \rightarrow j}$, drives the evolution towards consensus only when it reflects the tendency to copy a more successful neighbor. This mechanism may be compromised by large fluctuations in the difference of nodes' collected payoffs ($\pi_i -\pi_j$), when these payoffs come from just few interactions with a small number of neighbors. However, the enhanced final cooperation displayed by the HSBs is observed in general for all networks in Fig.~5(a-h), and specially for Metabolic and Music whose HSBs curves are visibly above the curves for the original networks (in black) for a wide range of initial conditions, $N_{\mathrm{coop}}^{\mathrm{ini}}/N_{\mathrm{gcc}} \in (0.4-0.9)$. For instance, the HSB with $\alpha^{\triangleright} =0.59$ of the Music network (see blue symbols curve in Fig.~5(h)) has 73$\%$ less links than the original network while preserving 83$\%$ of nodes, and still sustains up to $\approx 8$ times more final cooperation than the original network, for a fraction of initial cooperators of 0.8. To further ensure that the enhanced cooperation actually stems from the categorical structure of the backbones one should compare an HSB curve with that of its corresponding random surrogate. By doing so, we observe that the random surrogates happen to reproduce closer the cooperative behaviour of the original network, that is $LR$ curves in Fig.~5(a-h) follow the profile of the original network instead of appreciably deviating upwards, thus revealing that the surrogates do not provide a substantial gain in cooperation as opposite to HSBs. In general, the surrogates also require a higher proportion of initial cooperators, around $N_{\mathrm{coop}}^{\mathrm{ini}}/N_{\mathrm{gcc}} \gtrsim 0.6$, to produce any sizable increase in final cooperation with respect to the original network. This indicates that the internal hierarchical organization of the HSBs is key to sustain enhanced cooperation. Actually, the similarity filter preferentially removes links with lower hierarchy load, usually consisting of long-range connections stablished by high degree nodes, whereas the random removal makes no distinction. In fact, given the scale-freeness of real networks, deleting a long-range link at random is less likely due to their scarcity. Therefore, during the dynamics, similarity backbones may develop clusters of same-strategy nodes that are more stable through the evolutionary process than those found in random surrogates, the reason being the former are less exposed to distant contacts belonging to clusters of opposite strategy. This means the hierarchical structure of HSBs enables a better shielding for the groups of cooperators in the shape of metric clusters~\cite{Kleineberg2017metric}, which in turn can explain the increased cooperation levels found in similarity backbones.\\
To additionally validate our results we choose to explore four more different combinations of payoff values for the Music network in Appendix Fig.~\ref{Fig.11}. We observe that modifying the payoffs produces the same qualitative results as discussed above, with HSBs curves clearly surpassing the original network and evidencing that similarity backbones can reach superior cooperation.

\label{Conclusions}
\section{Discussion}
The existence of a metric space underlying complex networks allows us to provide an enriched interpretation of hierarchy that integrates two dimensions: popularity, or degree rank, and similarity between nodes, thus overcoming the problem of detecting hierarchies in the presence of clustering and the small world effect. The metric approach enables a unified framework to define the hierarchy loads of nodes and links.\\

Interestingly, the spectra of hierarchy loads of real networks revealed that, in general, these networks are less hierarchical than the reference provided by the maximum entropy null model and show greater variation in the hierarchy load of nodes across degree classes. Particularly, the lowest degree nodes typically contribute more towards the hierarchical structure, although their fluctuations are remarkable.

Moreover, we introduced the \textit{similarity filter}, a link pruning method which exploits the heterogeneity found in the hierarchy load of links. The filter extracts the connections that dominate the hierarchical structure of networks in terms of popularity and similarity, providing what we name hierarchical similarity backbones (HSB). The analysis of such backbones uncovered that, strikingly, the similarity filter is able to preserve network topological features at all scales while discarding a large number of links.   
Accordingly, from a fundamental point of view, hierarchical backbones could help provide new insight about the percolation properties of highly stratified real networks, aiding control of cascading failures, as well as have the potential to become a standard methodology for the detection, visualization and inspection of hierarchical clusters~\cite{Nielsen2016} in machine learning and data science environments.\\

From a practical point of view, the similarity filter has proven to be an exceptional tool to unravel the backbone that sustains enhanced cooperation in social dilemmas on structured populations. This is in line with previous simulations of prisoners dilemma type dynamics on adaptive networks, showing that cooperation combined with imitation can lead to a hierarchical structure~\cite{eguiluz2005cooperation}. When this dynamics is played on heterogeneous contact networks with underlying metric structure, the evolution of cooperation leads to the formation of clusters of cooperators in the similarity subspace~\cite{Kleineberg2017metric}. In the presence of these clusters, heterogeneity in the degrees was nevertheless found to hinder cooperation. Those findings reveal a tension between the popularity and similarity dimensions in evolutionary dynamics modelling social dilemmas. Our findings here solve this opposition by identifying the similarity backbones composed of significant links that are simultaneously hierarchical in terms of popularity and similarity, and which are expressly relevant in supporting and fostering cooperation.\\

Lastly, the methods developed in this contribution can be used to study the hierarchical nature of complex networks of any domain as long as they admit a geometric representation. The detection of hierarchical similarity backbones could for instance help in designing controllability of gene regulatory networks, improve communicability in information systems and infrastructures or assess robustness to species loss in ecological networks. Other possibilities for our framework include its extension to multiplex networks, opening promising future lines of research.

\setlength{\tabcolsep}{1.10em}
{\renewcommand{\arraystretch}{1.25}
\begin{table*}[]
\centering
    \begin{tabular}{l c c c c c c c c}
    \hline\hline
      \textbf{Data set}  & $N$   & $E$   &  $\beta$  & $k_{\mathrm{max}}$& $\av{k}$ & $\av{c}$ & $\av h$ & $\av h ^*$ \\ \hline
      Enron              & 182   & 2097  &  1.99  &     109           &  23.04   &  0.50  & 0.47 $\pm$ 0.19  &  0.77 $\pm$ 0.25 \\ \hline
      Internet           & 23752 & 58416 &  1.91  &     2778          &  4.92    &  0.61  & 0.29 $\pm$ 0.20  &  0.96 $\pm$ 0.10 \\ \hline
      Metabolic          & 1436  & 4718  &  2.15  &     224           &  6.57    &  0.54  & 0.45 $\pm$ 0.28  &  0.97 $\pm$ 0.09 \\ \hline
      Music				 & 2476  & 20624 &  2.30  &     1566		  &  16.66   &  0.82  & 0.34 $\pm$ 0.25  &  0.81 $\pm$ 0.23 \\ \hline\hline
    \end{tabular} 
    \caption{\textbf{Properties of the data sets under consideration}: $N$, size of the network; $E$ number of edges; parameter $\beta$ estimated from the embedding of the real network; $k_{\mathrm{max}}$, highest degree; $\av{k}$, average degree; and $\av{c}$, average clustering coefficient; $\av h$, average hierarchy load; $\av h ^*$, alternative average hierarchy load in terms of angular concentration.} \label{Tab:Real nets properties} 
\end{table*}
}

\label{Methods}
\section{Methods}

\label{Empirical data}
\subsection{Empirical data}

All real complex networks used in this paper have been mapped into their hyperbolic latent geometry using the embedding method Mercator~\cite{Garcia2019}. This method mixes machine learning and maximum likelihood approaches to infer the coordinates of the nodes in the underlying hyperbolic disk, while ensuring best congruency between the real network topology and the $\mathcal{S}^1$ geometric model. \\

\textbf{Enron.} The network captures the email communication  activity (125,409 emails) within employees from the Enron company. Edges are stablished between email addresses that shared correspondence. We use the dataset provided in \citep{Priebe2006} which includes also information about the organizational roles of 130 users.

\textbf{Internet.} We use the adjacency data for the Internet at the autonomous systems (AS) level assambled by the Archipelago project~\cite{Claffy2009} during June 2009.

\textbf{Human metabolic.} This network is the one-mode projection of metabolites of the bipartite metabolic network of human cell metabolisms. In this representation~\cite{Serrano2012}, there is a link between two metabolites if they participate in the same biochemical reaction.

\textbf{Music. }Nodes of the network are chords-sets of musical notes  (see \cite{Serra2012}) played in a single beat while edges represent detected transitions between these chords. We use a sparser, undirected version of the network reconstructed in \cite{Garcia2018Renorm}.

\label{Network geometric models}
\subsection{Models of Network Geometry}

\label{H2 model:}

\textbf{$\mathcal{H}^2$ model}\\
An isomorphism exists between the $\mathcal{S}^1$ and the $\mathcal{H}^2$ models~\cite{KrPa10}, so that hidden degrees $\kappa$ are mapped into radial coordinates, $r$, in a hyperbolic disk of radius $R_{\mathcal{H}^2}$, such that 
$  \kappa \sim e^{(R_{\mathcal{H}^2} \ - \ r)/2}$. Consequently, in the hyperbolic version, nodes with larger radial coordinates are located towards the edge of the hyperbolic disk and show lower expected degree. Particularly, in the $\mathcal{H}^2$ model every node $i$ is defined by the tuple $(r_i,\theta_i)$, and the probability that a link exists between two nodes $i$ and $j$ depends on their distance $d_{ij}$, as measured in the hyperbolic space using the hyperbolic law of cosines $ \mathrm{cosh}(d_{ij})=\mathrm{cosh}r_i\mathrm{cosh}r_j - \mathrm{sinh}r_i\mathrm{sinh}r_j\mathrm{cos}\Delta\theta_{ij}$. Nodes closely positioned in the hyperbolic disk have higher chances of being connected, thus the connection probability $p(d_{ij})$ must be a decreasing function of distance between them and, specifically, it can be chosen to be

\begin{equation}
\label{eq:conn_prob_hyp}
p(d_{ij}) = \frac{1}{1 + \exp\left[ \frac{\beta(d_{ij}-R_{\mathcal{H}^2})}{2}\right] },
\end{equation}

\noindent where the parameter $\beta$ still controls the network's clustering coefficient. In this paper, we mainly use the $\mathcal{S}^1$ model for calculation purposes, and its equivalent $\mathcal{H}^2$ version for visualization tasks.\\

\label{Soft Communities in Similarity Space (SCSS):}
\textbf{Soft Communities in Similarity Space (SCSS)}\\
The SCSS model~\cite{Garcia2018} produces synthetic geometric networks with inhomogeneous angular distributions, derived from geometric preferential attachement mechanisms, which were conceived in growing geometric network models~\cite{Zuev2015,Papadopoulos2012}. This means, for a network represented in an underlying (hyperbolic) metric space, the initial attractiveness of different angular regions during a geometric preferential attachemnt process is controlled by a parameter $\Lambda$. Consequently, this parameter regulates the heterogeneity of the angular coordinate, so that heteorgeneity is a decreasing function of $\Lambda$, with ${\rm{\Lambda }}\to \infty $ recovering the homogeneous distribution. The SCSS model, then takes such heterogeneous angular distribution defined by $\Lambda$ and adjusts it to an independent power-law degree distribution ($P(k)\sim k^{\gamma}$) and a tunable level of mean clustering $\av c$, controlled through parameter $\beta$. The SCSS model does so by introducing correlations between $\kappa$ and $\theta$ coordinates of nodes of the geometric network.\\

\label{Geometric Randomization (GR):}
\textbf{Geometric Randomization (GR)}\\ 
The GR~\cite{Starnini2019GR} is a model for the randomization of complex networks with geometric structure, which allows to uniformize their angular coordinate distribution, while preserving the exact degree sequence of the network. It thus applies to both real and synthetic networks where nodes have an observed degree and exist in a similarity space. In the GR model, angular coordinates $\theta$ are assigned to the nodes, chosen uniformly at random from [0, $2\pi$]. The network is then rewired following a likelihood maximization process that ensures the new topology is one generated by the ${{ \mathcal S }}^{1}$ model, while the observed degrees (and thus the number of edges) remain unaltered. The model is implemented using a single parameter $\beta$ controlling the mean clustering of the resultant rewired network. The rewiring and maximization procedure executed by the GR are specially useful to produce faithful real network replicas where only geometric (soft) communities have been supressed.

\subsection{Evolutionary Prisoner's dilemma}

The evolutionary prisoner's dilemma game~\cite{Szabo2007}, conducted on a network, considers that individual nodes playing with their contacts choose to either cooperate ($C$) or defect ($D$) every turn. The choice of strategies of the two interacting agents leads to specific payoffs, summarized by the payoff-matrix 

\setlength{\tabcolsep}{0.40em}
\begin{equation}
A=
\centering
\begin{tabular}{c|cc}
\centering
  & \textit{C} & \textit{D} \\ 
\hline 
\textit{C} & \textit{R} & \textit{S} \\
\textit{D} & \textit{T} & \textit{P}  
\label{PayoffMatrix}
\end{tabular} .
\end{equation}

\label{Evo.Game Theory}

That is, if both players cooperate, they both receive the \textit{reward} $R$ for cooperating. If both players defect, they both receive the \textit{punishment} payoff $P$. Lastly, if one of them defects while the other agent cooperates, the defector receives the \textit{temptation} payoff $T$, while the cooperator receives the "\textit{sucker}'s" payoff, $S$. In order for the game to be recognized as a prisoners's dilemma the condition $T > R > P > S$ must apply. Different ordinalities of the parameters define further classes of games~\cite{Nowak2006}
. In this paper, the prioner's dilemma is defined with parameter values: $T=1.5, R=1, P=0, S=-0.5$. Further parameter values are explored in Appendix. \\
The game proceeds in successive rounds. After each round, the strategies (C or D) of all nodes are updated synchronously according to the outcome of the imitation dynamics~\cite{Szabo2007,Cressman2014,Helbing1996} that outline the evolutionary mechanism. This means, during a single round, each individual node $i$ collects payoffs given by Eq.\ref{PayoffMatrix} from the interactions with all its neighbors and obtains an accumulated payoff $\pi_i$. All players chose then between their old strategy and the strategy of a randomly picked up neighbor $j$. In this way, node $i$ will adopt $j$'s strategy with a probability that depends upon the difference between collected payoffs of both nodes ($\pi_i - \pi_j$) as

\begin{equation}
p_{i\rightarrow j} =\frac{1}{1+e^{(\pi_i-\pi_j)/\tau}} \ , 
\end{equation}
which reflects the popular tendency of individuals to copy more successful neighbors. Such updating rule is a common choice in evolutionary dynamics~\cite{Poncela2012}, known as the Fermi rule since it is based in the Fermi distribution from Statistical Mechanics. The variable $\left(\frac{1}{\tau}\right)$, which in Physics stands for the inverse temperature, can be interpreted here as the intensity of the selection. That is, parameter $\left( \frac{1}{\tau}\right) $ (which we set to 0.5) controls the noise added to the decision-making process of, otherwise, perfectly rational players. After the simultaneous update of strategy of all nodes, their accumulated payoffs are reset and a new round begins.

\label{Prob.distribution}
\subsection{Probability distribution of angular distances between connected nodes in the $\mathcal{S}^1$ model}

The hierarchy load of an observed link, $h_{ij}$, can be computed analytically given the angular distance between the two nodes observed in the embedding, $\Delta \theta_{ij}^{\mathrm{obs}}$. To derive a closed expression for it, we first need to write down an expression for the probability distribution function of the angular separation between two nodes in the $\mathcal{S}^1$ model \textit{conditioned to the fact that they are connected}. Using Bayes' rule, we see that

\begin{equation}\label{eq:ang_dist_s1}
p \left( \Delta \theta_{ij} | a_{ij} = 1 \right) = \frac{ p \left( a_{ij} = 1 | \Delta \theta_{ij} \right) p \left( \Delta \theta_{ij} \right) }{ p \left( a_{ij} = 1 \right) },
\end{equation}

where $a_{ij}$ is the adjacency matrix element corresponding to the two nodes. In the above expression, $p \left( a_{ij} = 1 | \Delta \theta_{ij} \right)$ is the connection probability, Eq.~\eqref{eq:probability_of_connection_S1}, while the distribution of angular distances is simply $p \left( \Delta \theta_{ij} \right) = 1/\pi$, given that angular coordinates are homogeneously distributed in the $\mathcal{S}^1$ model. The denominator can be obtained by direct integration, $p \left( a_{ij} = 1 \right) = \int_0^{\pi} p \left( a_{ij} = 1 | \Delta \theta_{ij} \right) p \left( \Delta \theta_{ij} \right) \mathrm{d} \Delta \theta_{ij}$. Using the definition of $h_{ij}$, we obtain

\begin{equation}\label{eq:h_ij_explicit}
\begin{split}
\begin{aligned}
h_{ij} &= P \left( \Delta \theta_{ij} > \Delta \theta_{ij}^{\mathrm{obs}} \right) \\
&= 1 - \int \limits_0^{\Delta \theta_{ij}^{\mathrm{obs}}} p \left( \Delta \theta_{ij} | a_{ij} = 1 \right) \mathrm{d} \Delta \theta_{ij} \\
&= 1 - \frac{\int \limits_0^{\Delta \theta_{ij}^{\mathrm{obs}}} p \left( a_{ij} = 1 | \Delta \theta_{ij} \right) p \left( \Delta \theta_{ij} \right) \mathrm{d} \Delta \theta_{ij} }{ \int \limits_0^{\pi} p \left( a_{ij} = 1 | \Delta \theta_{ij} \right) p \left( \Delta \theta_{ij} \right) \mathrm{d} \Delta \theta_{ij} } \\
&= 1 - \frac{\Delta \theta_{ij}^{\mathrm{obs}} {}_2 F_1 \left( 1, \frac{1}{\beta}; 1 + \frac{1}{\beta}; - \left( \frac{R \Delta \theta_{ij}^{\mathrm{obs}}}{\mu \kappa_i^{\mathrm{obs}} \kappa_j^{\mathrm{obs}}} \right)^{\beta} \right)}{\pi {}_2 F_1 \left( 1, \frac{1}{\beta}; 1 + \frac{1}{\beta}; - \left( \frac{R \pi}{\mu \kappa_i^{\mathrm{obs}} \kappa_j^{\mathrm{obs}}} \right)^{\beta} \right)},
\end{aligned}
\end{split}
\end{equation}

where ${}_2 F_1 (a, b; c; z)$ is the hypergeometric function. Equation \eqref{eq:h_ij_explicit} yields the hierarchy load of a given link in terms of the angular separation and product of hidden degrees of the nodes at both ends of the edge, as well as of the global parameters $R$, $\mu$, and $\beta$. 

Finally, let us also show that the expected value of $h_{ij}$ for any link in a network generated by the $\mathcal{S}^{1}$ model is $\langle h_{ij} \rangle_{\mathcal{S}^{1}} = 1/2$. To calculate $\langle h_{ij} \rangle_{\mathcal{S}^{1}}$, we only need to notice that, in this case, the angular separation between the nodes at the ends of the edge in the resulting network, $\Delta \theta_{ij}^{\mathrm{obs}}$, is itself a random variable with distribution $\rho \left( \Delta \theta_{ij}^{\mathrm{obs}} \right)$ given by Eq.~\eqref{eq:ang_dist_s1} (that is, $\rho \left( \Delta \theta_{ij}^{\mathrm{obs}} \right) = p \left( \Delta \theta_{ij}^{\mathrm{obs}} | a_{ij} = 1 \right)$ and, therefore, the expected value of $h_{ij}$ is

\begin{equation}
\begin{split}
\begin{aligned}
\langle h_{ij} \rangle_{\mathcal{S}^{1}} &= \left\langle P \left( \Delta \theta_{ij} > \Delta \theta_{ij}^{\mathrm{obs}} \right) \right\rangle_{\mathcal{S}^{1}} \\
&= \int \limits_{0}^{\pi} \rho \left( \Delta \theta_{ij}^{\mathrm{obs}} \right) P \left( \Delta \theta_{ij} > \Delta \theta_{ij}^{\mathrm{obs}} \right) \mathrm{d} \Delta \theta_{ij}^{\mathrm{obs}}\\
&= \int \limits_{0}^{\pi} \rho \left( \Delta \theta_{ij}^{\mathrm{obs}} \right) \int \limits_{\Delta \theta_{ij}^{\mathrm{obs}}}^{\pi} \rho \left( \Delta \theta_{ij} \right) \mathrm{d} \Delta \theta_{ij} \mathrm{d} \Delta \theta_{ij}^{\mathrm{obs}} = \frac{1}{2}.
\end{aligned}
\end{split}
\end{equation}
In the last step, we have used the fact that $\rho(z)$ is normalized, $\int_{0}^{\pi} \rho \left( z \right) \mathrm{d}z = 1$.

\vspace*{4cm}
\bibliographystyle{naturemag}
\bibliography{ref_full.bib}

\label{Acknowledgements}
\section{Acknowledgements}
M.\'{A}.S. acknowledge support from a James S McDonnell Foundation Scholar Award in Complex Systems; Ministerio de Ciencia, Innovación y Universidades of Spain project no. FIS2016-76830-C2-2-P (AEI/FEDER, UE); the project Mapping Big Data Systems: embedding large complex networks in low-dimensional hidden metric spaces—Ayudas Fundación BBVA a Equipos de Investigación Científica 2017, and the Generalitat de Catalunya grant No. 2017SGR106. G.G.-P.~acknowledges financial support from the Academy of Finland via the Centre of Excellence program (Project no.~312058 as well as Project no.~287750), and from the emmy.network foundation under the aegis of the Fondation de Luxembourg.

\section{Author Contributions}
E.O., G.G.-P. and M.\'{A}.S. contributed to the design and implementation of the research, to the analysis of the results and to the writing of the manuscript.

\section{Competing Interests statement}
The authors declare no competing financial interests.

\section{Data availability}
Codes and data supporting the findings of this study are available from the corresponding author upon request.

\section{APPENDIX}

\subsection{Hierarchy load spectrums of nodes for GR network replicas}

\begin{figure}[b]
\centering
  \includegraphics[width=1.0\linewidth]{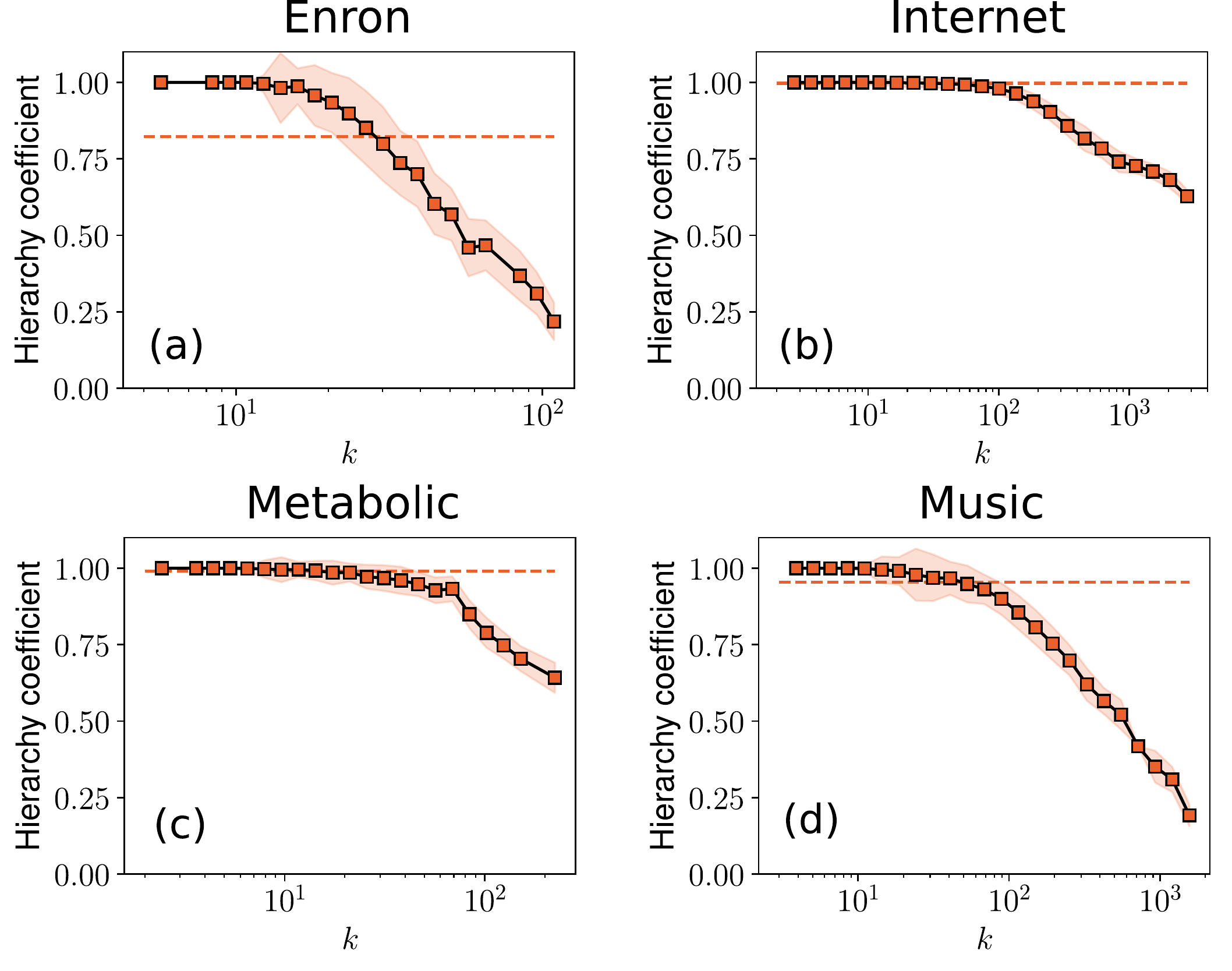}
  \caption{\textbf{Hierarchy load spectrums} for Geometric Randomization (GR) model replicas of each of the 4 real networks under study. Each spectrum is the average over 10 independent realizations. Dashed lines indicate the average hierarchy load of the network, obtainded as $\av h^* = N^{-1}\sum _{i=1}^{N} h_i^*$, yielding to the following values for each network: $\av h ^*_{\mathrm{Enron}} = 0.82 \pm 0.19$, $\av h ^*_{\mathrm{Internet}} = 0.99 \pm 0.02$, $\av h ^*_{\mathrm{Metabolic}} = 0.99 \pm 0.04$ and $\av h^*_{\mathrm{Music}} = 0.95 \pm 0.11$. }
  \label{Fig.6}
\end{figure}

\begin{figure*}[t]
\centering
  \includegraphics[width=0.93\linewidth]{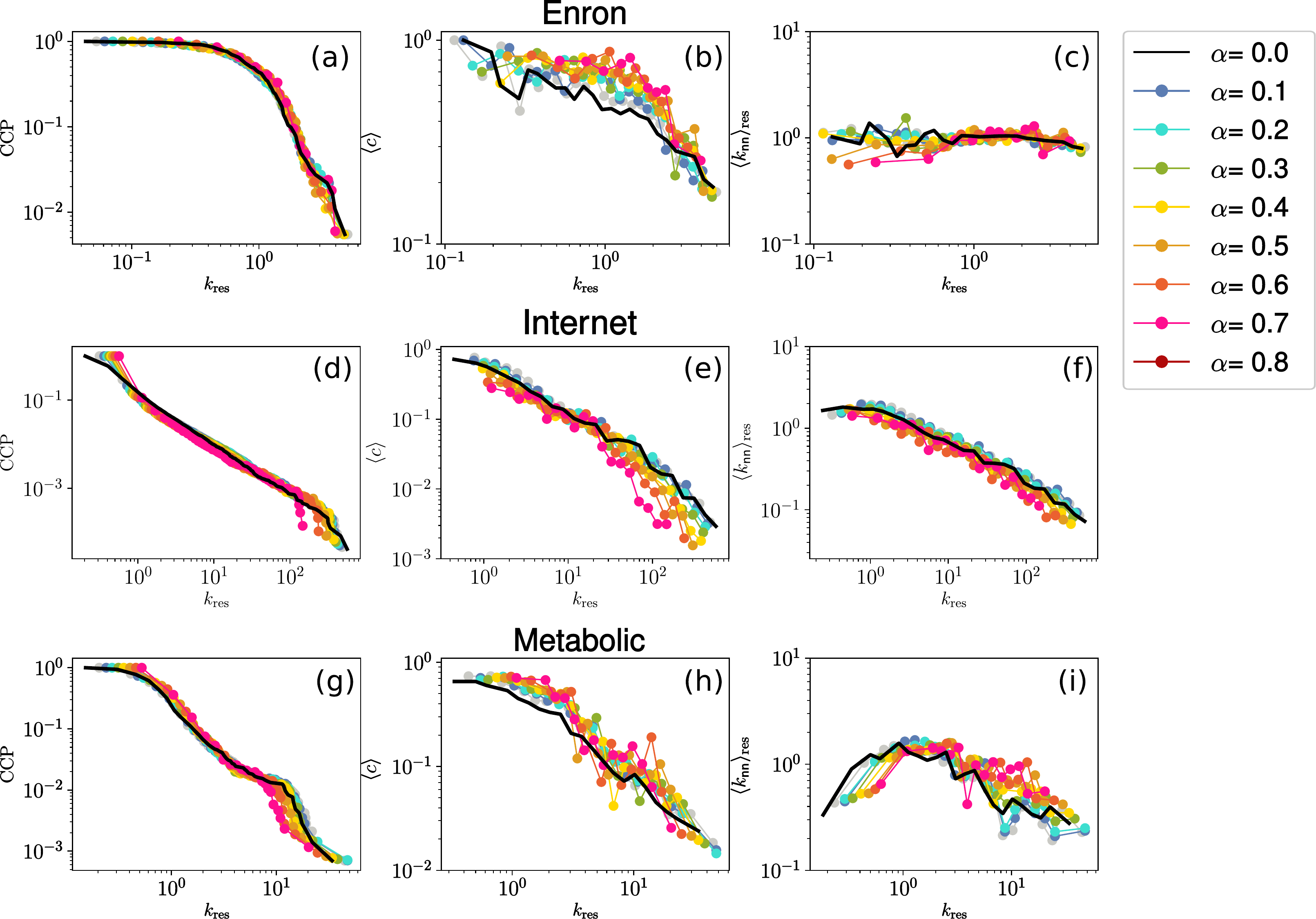}
  \caption{\textbf{Topological features of the hierarchical backbones} of (a-c) Enron, (d-f) Internet, (g-i) Metabolic ,  obtained for $\alpha \in [0.1-0.8]$.  In each case, value $\alpha=0.0$ corresponds to the original network. The properties in each row, from left to right, are: complementary cummulative distribution of rescaled degrees ($k_{\mathrm{res}} = k/\langle k\rangle $), degree dependent clustering coefficient over rescaled-degree classes, normalised average nearest neighbor degree $\langle k_{\mathrm{nn}} \rangle_{\mathrm{res}}= \langle k_{\mathrm{nn}}(k_{\mathrm{res}})\rangle\langle k \rangle /\langle  k^2 \rangle$.  
  }
  \label{Fig.7}
\end{figure*}

Figure~2(c-f) shows the spectrum of hierarchy loads of nodes in terms of angular concentration, $h_i^*$, for 4 real networks. Here, we provide the same metric for angularly randomized versions of such real networks, obtained using the Geometric Randomization model (see Methods~B in main paper). By comparing the spectra in Fig.2(c-f) and Fig.~\ref{Fig.6} in this Appendix one can notice that the complete homogenization of the angular coordinates of nodes (and thus the full elimination of geometric communities) does not  translate into radical changes in the hierarchy load profiles. Instead, below (Fig.~\ref{Fig.6}) we observe for each of the randomized networks that the inverse correlation with the degree of the hierarchy loads is compatible with that of the original network, and that the average hierarchy load $\av h^*$ values of the GR networks are not remarkably different to that of the real ones. The most noticeable influence of geometric community structure in the hierarchy load measurement is found for the Metabolic network, which turns to be a natural effect 
 given the pronounced bimodal distribution of node angles of the original real network.

\setlength{\tabcolsep}{2.10em}
{\renewcommand{\arraystretch}{1.0}
\begin{table*}[t]
\centering
\begin{tabular}{lrrrr}
\hline
\textbf{Network} & \textbf{Filtering method} & \textbf{Threshold} & \textbf{N$_{\mathrm{gcc}}$}   & \textbf{E$_{\mathrm{gcc}}$}   \\\hline\hline
Enron & none     & --      & 182   & 2097   \\
Enron & HSB     & $\alpha^{o}$=0.66      & 159    & 587    \\
Enron & HSB     & $\alpha^{\rhd}$=0.68      & 136    & 498    \\ 
Enron & RLR    & LR$^{o}$=1510   & 175.30 $\pm$ 1.42 & 587 \\
Enron & RLR    & LR$^{\rhd}$=1599   & 174.30  $\pm$ 2.45 & 498 \\
\hline
Internet & none     & --      & 23748 & 58414   \\
Internet & HSB     & $\alpha^{o}$=0.18      & 18536 & 27374\\
Internet & HSB     & $\alpha^{\rhd}$=0.27      & 16691 & 22970\\ 
Internet & RLR    & LR$^{o}$=31040      & 15901.20 $\pm$ 36.06      & 27374 \\
Internet & RLR    & LR$^{\rhd}$=35444      & 14172.80 $\pm$ 40.27      & 22970 \\\hline
Metabolic & none     & --      & 1436  & 4718     \\
Metabolic & HSB     & $\alpha^{o}$=0.21      & 1157 & 2717\\
Metabolic & HSB     & $\alpha^{\rhd}$=0.25      & 1069 & 2470\\ 
Metabolic & RLR    & LR$^{o}$=2001      & 1195.60 $\pm$ 14.21   & 2717 \\
Metabolic & RLR    & LR$^{\rhd}$=2248      & 1147.10 $\pm$ 13.35  & 2470 \\\hline
Music & none     & --      & 2476  & 20624    \\
Music & HSB     & $\alpha^{o}$=0.53      & 2252 & 6617\\
Music & HSB     & $\alpha^{\rhd}$=0.59      & 2111 & 5774\\ 
Music & RLR    & LR$^{o}$=14007      & 1905.60 $\pm$ 12.04    & 6617 \\
Music & RLR    & LR$^{\rhd}$=14850      & 1824.60 $\pm$ 21.49    & 5774 \\ \hline \\

\end{tabular}
\caption{\textbf{Backbones used to study evolutionary game dynamics.} HSB stands for hierarchical similarity backbone, meaning these backbones are obtained using the similarity filter with specific $\alpha$. RLR stands for random link removal, thus this method gives random surrogates where a specific number of links have been removed, indicated by LR (links removed). The results in Fig.~5(i-l) and Appendix Fig.~10 are the average of 10 random surrogate realizations. Because of this, and since by construction we fix the number of edges in the random surrogates, only their number of nodes can fluctuate and hence display an error interval.}
\label{Tab:Backones_Evo_DYn}
\end{table*}}

\subsection{Topological properties of similarity backbones of real networks}

First, in Fig.~7 we provide results for the topological features of similarity backbones examined in Fig.~4(g-i) for the rest of real networks under study. \\
Secondly, in Fig.~8 we cover further topological metrics of the HSBs and showcase them against the $\alpha$ parameter controlling the filtering procedure. 

\subsection{Composition of similarity backbones of real networks}

In this section we sort the nodes in the network from highest degree to lowest and tag as "hubs" all nodes lying within a top slice of the list,
delimited by a threshold value $\tau$. Subsequently, in Fig.~9 we keep track of the proportion of such high-degree nodes in
every similarity backbone of increasing $\alpha$ for the 4 real networks. The results are analogous to those of Fig.~4(f) but obtained for an extended range of $\tau$ values.

\subsection{Evolutionary Prisoner's dilemma game dynamics}
First, for the four real networks analysed, we report in Table S1  the number of nodes and edges in the $gcc$s of every similarity backbone and random surrogate along with $\alpha$ filtering values and number of links removed, respectively. Secondly, in Fig.~10 we provide analogous plots of those in Fig.~4(g-i) (main paper) and Fig.~7 in Appendix for the random surrogates, showcasing their main topological features. Lastly, in Fig.~11, we explore the dynamics on similarity backbones (HSB) for further payoff values for one of the networks (Music).

\begin{figure*}[t]
\centering
  \includegraphics[width=0.83\linewidth]{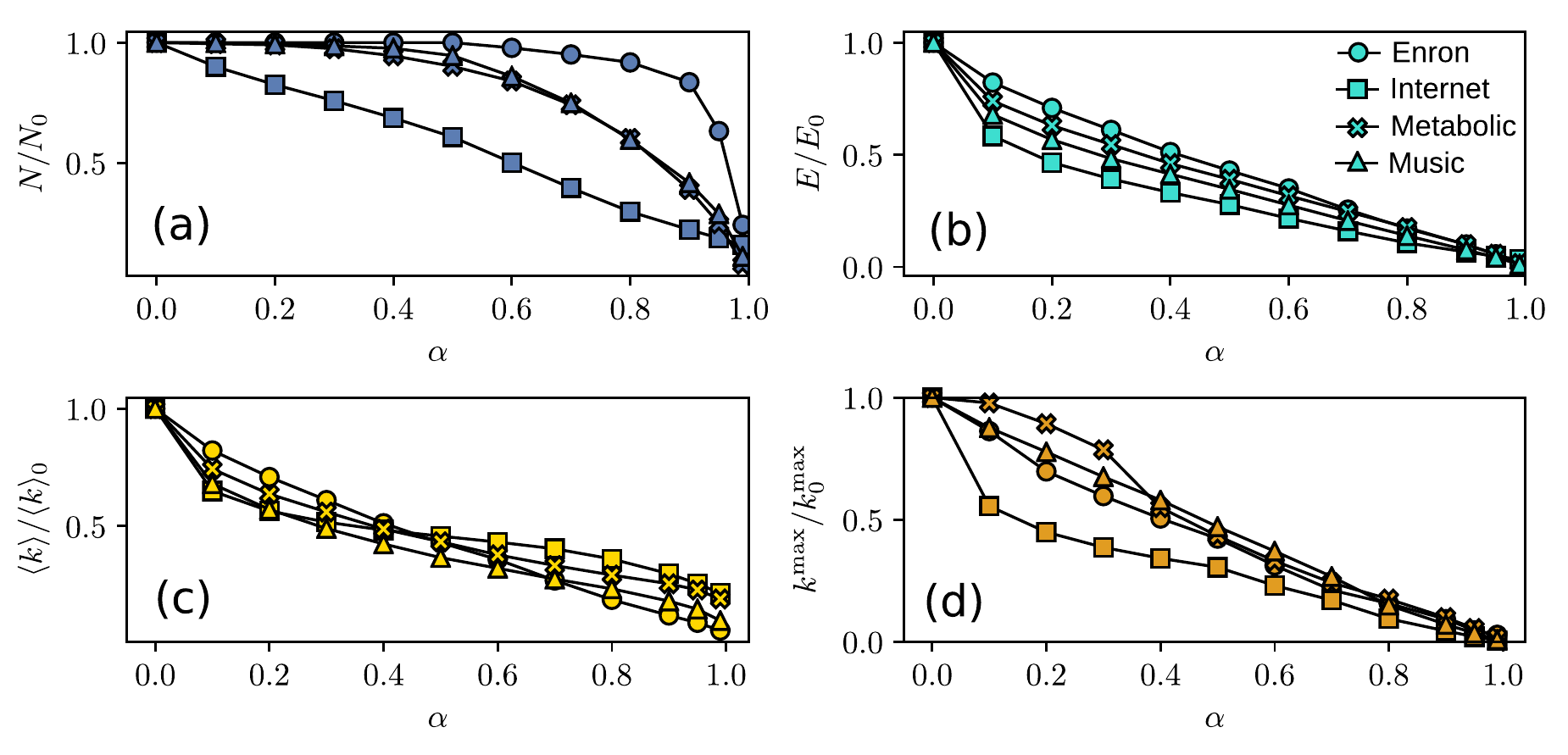}
  \caption{\textbf{Topological properties of hierarchical backbones} of the four real networks against increasingly restrictive filtering parameter ($\alpha$) values. Plots from (a) to (d) show, for increasing values of $\alpha$: the normalised number of nodes in the backbone; the normalised number of edges in the backbone; normalised average degree of the backbone and the HSB normalised maximum degree. }
  \label{Fig.8}
\end{figure*}

\begin{figure*}[t]
\centering
  \includegraphics[width=0.90\linewidth]{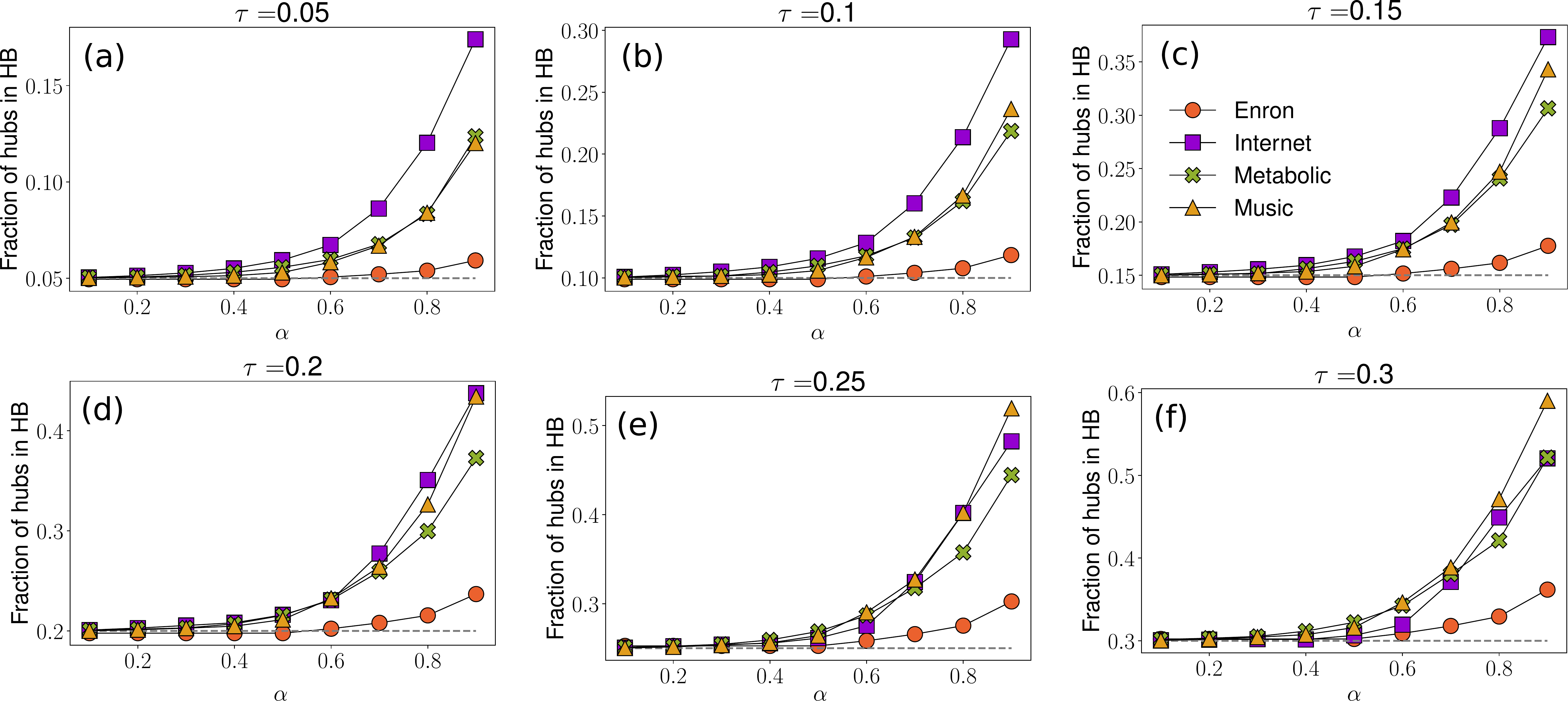}
  \caption{\textbf{Fraction of nodes considered hubs for threshold values in range $\tau = [0.05 - 0.30]$}, found in backbones obtained with increasing $\alpha$ for the real networks Enron, Internet, Metabolic and Music.}
  \label{Fig.9}
\end{figure*}

\begin{figure*}[t]
\centering
  \includegraphics[width=0.83\linewidth]{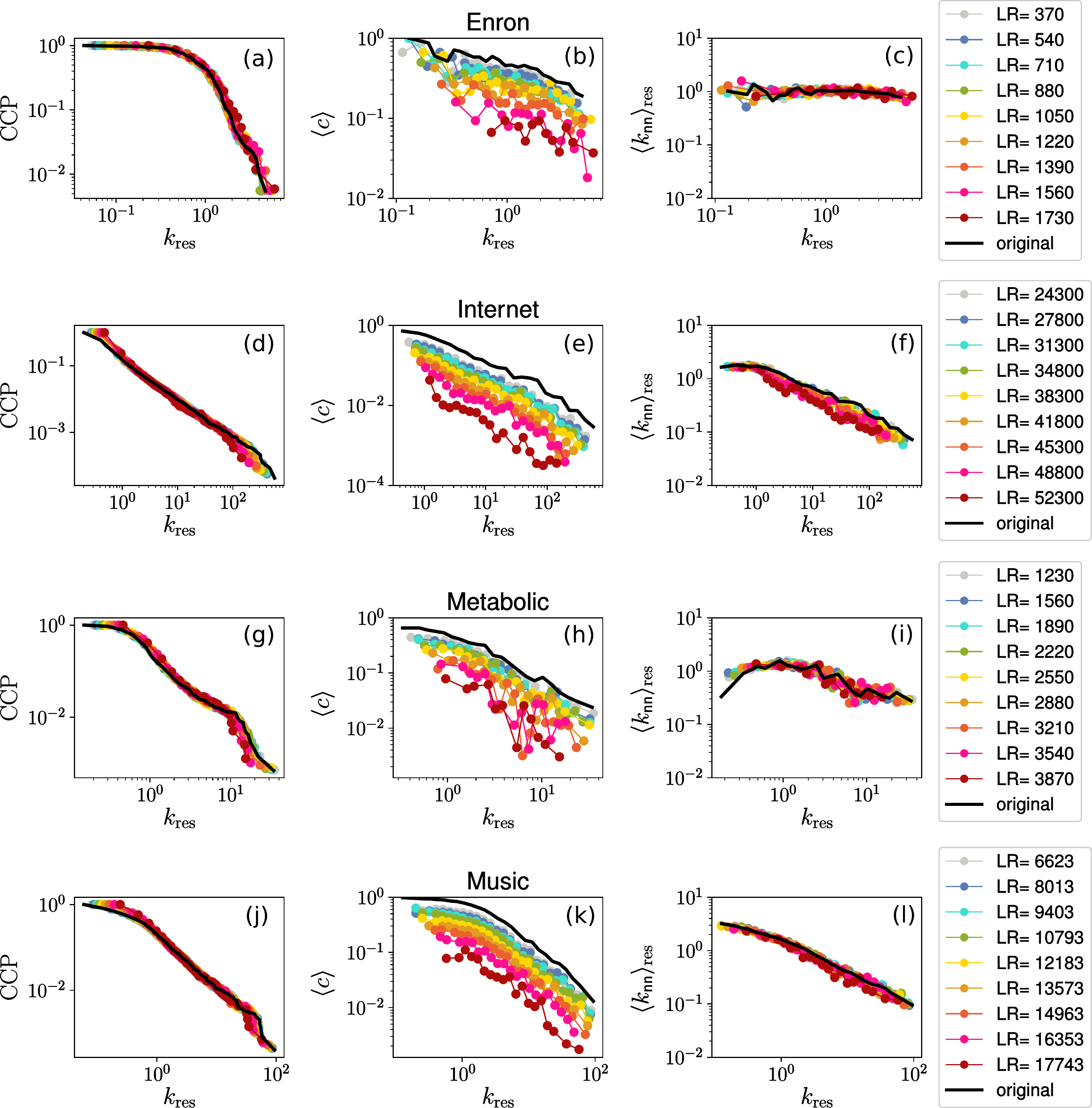}
  \label{Fig.10}
\end{figure*}

\begin{figure*} [t!]
   \caption{\textbf{Topological features of random surrogates} and original network of (a-c) Enron, (d-f) Internet, (g-i) Metabolic (j-l) Music.  In each case, value $LR$ indicates the number of links that have been removed at random to produce the surrogate backbone. The properties in each row, from left to right, are: complementary cummulative distribution of rescaled degrees $k_{\mathrm{res}} = k/\langle k\rangle $, degree dependent clustering coefficient over rescaled-degree classes, normalised average nearest neighbor degree $\langle k_{\mathrm{nn}} \rangle_{\mathrm{res}}= \langle k_{\mathrm{nn}}(k_{\mathrm{res}})\rangle\langle k \rangle /\langle  k^2 \rangle$. \vspace*{1.0cm}
   }
     \includegraphics[width=1.0\linewidth]{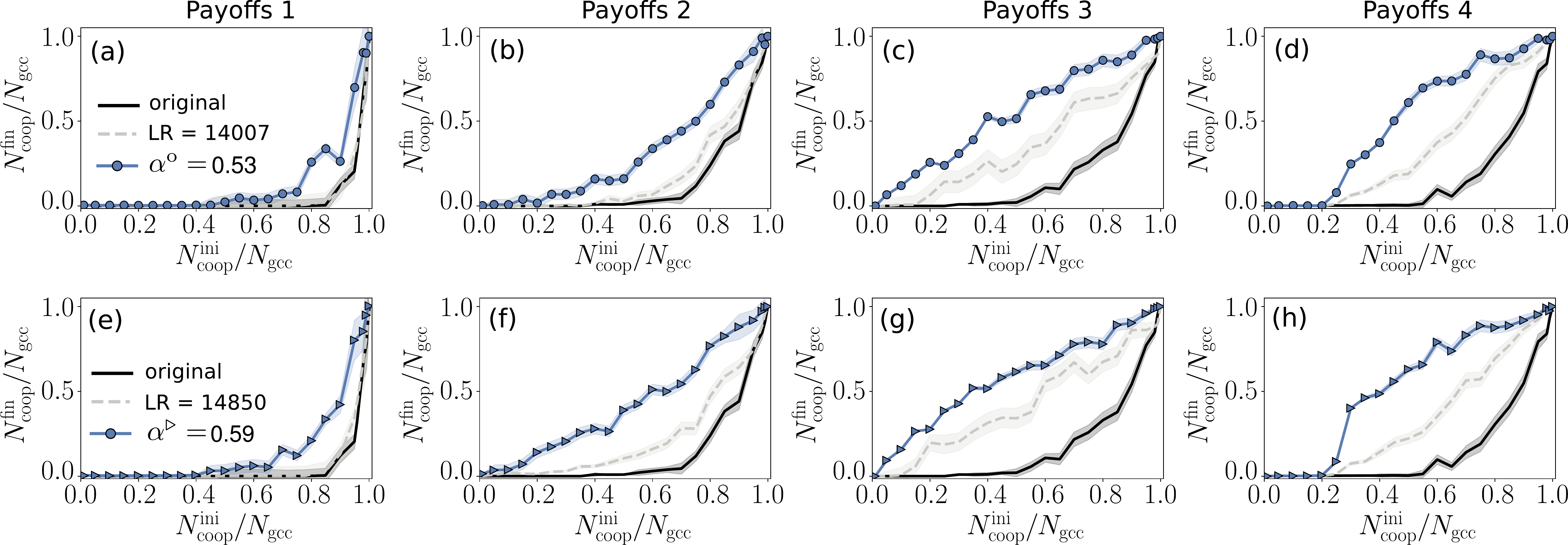}
       \caption{ \textbf{Fraction of final cooperators against fraction of initial cooperators for similarity backbones and random surrogates of Music network under different payoffs}. The results of the dynamics in each plot are obtained using the following payoff values for the evolutionary prisoner's dilemma game: \textbf{(a) and (e)} $T=1.0, \ R=0.5, \ P=-0.5, \  S=-1.0$ \textbf{(b) and (f)} $T=1.5, \  R=1.0, \  P=0.5, \  S=0.0$ \textbf{(c) and (g)} $T=2.0, \  R=1.5, \  P=0.5, \  S=0.0$ \textbf{(d) and (h)} $T=2.0, \  R=1.5, \  P=0.0, \  S=-0.5$. Results for the original network appear for reference, indicated by a black line. HSB filtered with $\alpha= 0.53$ are shown as blue curves with circles whereas HSBs filterd with $\alpha=0.59$ correspond to blue curves with triangles in the second row. Results of random surrogates are displayed as grey dashed lines.}
       
  \label{Fig.11}
\end{figure*}

\end{document}